# Magnetic structure of the field-induced multiferroic $GdFe_3(BO_3)_4$

Haiding Mo,[1] Christie S. Nelson,[1] Leonard N. Bezmaternykh,[2] and V.T. Temerov[2]

[1]*National Synchrotron Light Source, Brookhaven National Laboratory, Upton, NY 11973, USA*

[2]*L. V. Kirensky Institution of Physics. Siberian Branch of RAS, Krasnoyarsk, 660036, Russia*

Abstract

We report a magnetic x-ray scattering study of the field-induced multiferroic $GdFe_3(BO_3)_4$. Resonant x-ray magnetic scattering at the Gd $L_{II, III}$ edges indicates that the Gd moments order at $T_N \approx 37$ K. The magnetic structure is incommensurate below $T_N$, with the incommensurability decreasing monotonically with decreasing temperature until a transition to a commensurate magnetic phase is observed at $T \approx 10$ K. Both the Gd and Fe moments undergo a spin reorientation transition at $T_{SR} \approx 9$ K such that the moments are oriented along the crystallographic *c* axis at low temperatures. With magnetic field applied along the *a* axis, our measurements suggest that the field-induced polarization phase has a commensurate magnetic structure with Gd moments rotated ~45° toward the basal plane, which is similar to the magnetic structure of the Gd subsystem observed in zero field between 9 and 10 K, and the Fe subsystem has a ferromagnetic component in the basal plane.

# I. INTRODUCTION

Studying the magnetic structure and phase transitions in crystals that have two or more magnetic subsystems is of fundamental importance for understanding the mutual interaction between different spins and their couplings to other degrees of freedom. Unsurprisingly, these couplings give rise to complex phase diagrams and a wide variety of phase transitions driven by temperature or magnetic field, such as paramagnetic-to-antiferromagnetic, spin reorientation, spin flop, and commensurate-to-incommensurate. One example of these materials is the family of rare-earth ferroborates, $RFe_3(BO_3)_4$, which have received much attention recently. They crystallize in the huntite-type structure with space group *R*32 at room temperature. Both the rare earth and iron ions carry magnetic moments, and these materials exhibit interesting optical and magnetic properties.[1-4]

The rich phase diagram of the rare-earth ferroborate $GdFe_3(BO_3)_4$ has been revealed by measurements of the dielectric constant,[3] magnetization and magnetic susceptibility,[3,5-7] Mössbauer spectroscopy,[7] and antiferromagnetic resonance.[8,9] Two magnetic transitions have been inferred from these studies in addition to a structural one at 156 K. At $T_N \approx 37$ K, the crystal undergoes a phase transition from a paramagnetic to an antiferromagnetic phase, with the Fe moments thought to lie in or close to the *ab* basal plane, arranged ferromagnetically in the basal plane, and stacking antiferromagnetically along the crystallographic *c* axis. At $T_{SR} \approx 9$ K, the spin reorientation temperature, the easy axis is thought to rotate $90^o$ with the Fe moments oriented along the *c* axis, forming a new antiferromagnetic phase. However, the detailed magnetic structure in each phase is unclear. Mössbauer spectroscopy indicates that even in zero magnetic field, the moments are actually tilted away from these assumed directions.[7] More importantly, the models used so far to explain the behavior of $GdFe_3(BO_3)_4$ assume that the Gd moments remain in a paramagnetic state below $T_N$,[2,4] but antiferromagnetic resonance measurements indicate that Gd orders simultaneously with the Fe subsystem at $T_N$.[9] In addition, dielectric constant measurements suggest that there is a phase transition at $T_M$, which is 0.7 K above the spin reorientation temperature in zero magnetic field. The nature of the transition at $T_M$ is unknown.

Other than these interesting magnetic properties, recent studies have shown that Gd and Nd ferroborates are field-induced multiferroics.[2,10] Currently, there is great interest in multiferroic materials because of the fundamental aspects of the mechanism that gives rise to the magnetic-ferroelectric coupling, and their promise of technological applications.[11] In the last several years, many theories have been proposed to explain the origin of electric polarization in magnetically ordered materials,[12-15] but questions remain about the relationship between ferroelectricity and the magnetic structure.[14,15] In Gd ferroborate, field-induced polarization (FIP) is observed for $T_{SR} < T < T_M$, and this temperature range depends upon the direction of the applied field. For $\mathbf{H}||\boldsymbol{c}$, $T_M$ is unchanged while $T_{SR}$ decreases,[3] so FIP is only observed for T < 9.7 K. For low field $\mathbf{H}||\boldsymbol{a}$, $T_M$ increases while $T_{SR}$ is unchanged.[3] A stronger $\mathbf{H}||\boldsymbol{a}$ results in a decrease in $T_{SR}$, and therefore FIP is observed for $T < T_N$. Regardless of the field direction, FIP in Gd ferroborate occurs only when the moments are close to the basal plane.[2] However, because the detailed magnetic structure in each phase is unknown, it is difficult to determine what mechanism is responsible for the multiferroic behavior. A study of the magnetic structure and phase transitions in Gd ferroborate is needed to shed light on the origin of its multiferroic behavior.

X-ray and neutron scattering are direct probes of magnetic order, and can provide information about the order parameter, correlation length, and moment orientation. Neutron scattering has been used to study the magnetic structure of Nd, Tb, Ho, and Y ferroborates, and the results have confirmed that the compounds have antiferromagnetic order below $T_N$ with the propagation vector of (0 0 3/2).[16-18] The moment orientation has been found to vary with rare earth or Y ion, ranging from the basal plane (Nd and Y)[16,18] to the c axis (Tb)[17], and with non-coplanar subsystems and a spin reorientation at $T_{SR}$ = 5 K in Ho ferroborate.[18] However, it is difficult to study Gd ferroborate with neutron scattering because of the large neutron absorption cross-section of gadolinium. X-ray scattering provides an alternative for characterization of the magnetic structure of this material. In addition, one can tune the x-ray energy to the absorption edges, and use resonant x-ray scattering to gain element specificity. In particular, x-ray resonant scattering at the Gd $L_{II,III}$ edges enables one to probe the order in the Gd magnetic subsystem, thus determining directly whether or not the Gd moments order in each phase.

Previous x-ray resonant scattering studies of magnetic structure in Gd compounds have reported that there is a large resonant enhancement of magnetic peak intensity when the incident x-ray energy is tuned to the Gd $L_{II}$ or $L_{III}$ edges if the Gd moments are ordered in the crystal.[19-26]

In this paper, we report an x-ray magnetic scattering study of $GdFe_3(BO_3)_4$. Our resonant scattering data indicate that Gd moments order at $T_N$. We also verify the spin reorientation transition at $T_{SR} \approx 9$ K in zero field: the Gd and Fe moments reorient toward the *c* axis below $T_{SR}$. Approximately 1 K above $T_{SR}$, there is a phase transition from a commensurate magnetic (CM) phase to a long period, incommensurate magnetic (ICM) phase as manifested in the splitting of the magnetic peaks, observed at a wavevector of (0 0 3/2 ± ε), where ε is the incommensurability. Above 10 K, ε increases continuously as a function of temperature and reaches ~0.002 $c^*$ near $T_N$. With a magnetic field applied along the *a* axis, the reorientation temperature decreases and the FIP phase appears to have a CM structure.

## II. EXPERIMENTAL METHODS

Single crystals of $GdFe_3(BO_3)_4$ were grown as described in reference 6. The crystal used in this experiment is ~6×6×3 mm$^3$, with naturally-formed, smooth, triangular (001) surfaces. The mosaic spread of the (001) surface is 0.02$^o$ as determined by the full width at half maximum of the (003) structural Bragg peak. X-ray scattering measurements with and without magnetic field were conducted at beamlines X21 and X22C, respectively, at the National Synchrotron Light Source. Both beamlines employ Si(111) double crystal monochromators, and have an energy resolution of ~5 eV at the Gd $L_{II}$ edge (7.93 keV).

At beamline X22C, the sample was mounted with the (0 0 1) face in a vertical scattering geometry, inside a closed-cycle cryo-refrigerator from Advanced Research Systems, Inc. The lowest temperature of the cooling system is ~1.5 K. In this scattering configuration, which is shown in the inset to Fig. 1b, the incident beam is primarily σ polarized with a polarization of ~90% as determined by measuring the incident beam and the (003) and (006) structural Bragg peaks, which agrees well with previous measurements.[27] For temperature- and azimuth-dependent measurements, a PG(002) crystal with a mosaic width of ~0.3° was used as the analyzer to reduce the background; for high resolution scattering measurements, a Ge(111) crystal was used as an analyzer; and for polarization analysis, MgO(004) (mosaic width ~0.1°) and Cu(220) (mosaic width ~0.3°) crystals were used as analyzers for resonant scattering at the Gd $L_{II}$ edge and for nonresonant scattering at 7 keV, respectively. At the Gd $L_{II}$ edge, the scattering angle of MgO(004) is 95.7°, and at 7 keV, the scattering angle of Cu(220) is 87.7°, which leads to ~1% of leakage in the intensity for resonant scattering at the Gd $L_{II}$ edge, and ~0.2% for nonresonant scattering at 7 keV.

At beamline X21, x-ray scattering in a magnetic field was carried out with a vertical magnetic field and a horizontal scattering plane (see inset of Fig. 5a). The applied magnetic field is along the *a* axis and perpendicular to the $b^*c$ scattering plane. The (0 0 9/2) magnetic peak intensity was measured at the Gd $L_{II}$ edge and at 7 keV, as functions of both field and temperature. In order to reduce the background, a PG(002) crystal was used as an analyzer with the scattering plane of the analyzer also horizontal. The field-dependent data were taken from low to high field at a fixed temperature, and the temperature-dependent data were taken from low to high temperature at a fixed magnetic field, all after zero-field cooling.

## III. RESULTS

### A. ZERO-FIELD RESONANT X-RAY SCATTERING

Below $T_N$, additional peaks are found at (h k l) ± (0 0 3/2), where (h k l) is a structural Bragg peak satisfying –h + k + l = 3n for n integer, which is the extinction rule of the high temperature *R*32 phase. This suggests that the magnetic unit cell is doubled along the *c* axis.

Energy scans near the Gd $L_{II,III}$ edges exhibit large resonant enhancements at the (0 0 3/2) wavevector, as shown in Fig. 1 for the (0 0 3/2) peak. This can be compared to the reduction in peak intensity observed at the (0 0 3) structural Bragg peak, which is due to the increase in absorption. The peak positions in the energy scans at the (0 0 3/2) wavevector are slightly above the absorption edges, which indicates the dipole character of the electronic transitions. To further verify the magnetic origin of the peaks, we used polarization analysis at the (0 0 9/2) peak, and the results are shown in the inset of Fig. 1a. As expected for resonant x-ray magnetic scattering, the $\pi$ component dominates, and the $\sigma$ component, which is ~10% of the total intensity, can be attributed to the $\pi$ component of the incident beam. The polarization dependence is therefore consistent with a magnetic origin of the peak.

The large resonant enhancement at the Gd $L_{II}$ edge at the magnetic wavevector results in a resonant signal that is at least 100 times larger than that from the nonresonant scattering. Therefore the latter can be ignored, and the resonant scattering can be used to study the magnetism of the Gd subsystem. Near T = 2 K, the ratio of the (0 0 9/2) magnetic peak intensity to the (0 0 6) structural Bragg peak intensity is $\sim 2.5 \times 10^{-4}$. Note that this is about 3 orders of magnitude larger than that observed in $GdFe_4Al_8$, which is a compound that also has both Gd and Fe magnetic subsystems. In $GdFe_4Al_8$ the resonant scattering was attributed to the polarization of the d-band but not the ordering of the Gd 4f moments,[28] and therefore the much larger enhancement observed in $GdFe_3(BO_3)_4$ is clear evidence of the ordering of its 4f moments.

In Fig. 2, the temperature dependence of the (0 0 9/2) magnetic peak measured at the Gd $L_{II}$ edge is shown. The integrated intensity is observed to decrease with increasing temperature until vanishing at $T_N \approx 37$ K, which is further confirmation of its magnetic origin. An energy scan near T = 36 K (not shown) verifies that the intensity is still dominated by the resonant scattering. As shown in the inset to Fig. 2a, the data close to $T_N$ were fitted using a power law, $I = A(T_N - T)^{2\beta}$, to determine an accurate value of $T_N$.[25] The fit gives $T_N = 36.5(\pm 0.1)$ K and $\beta = 0.596(\pm 0.01)$. $T_N$ is consistent with the values obtained from previous measurements.[3,8]

The width of the magnetic peak was observed to increase at T ≈ 10 K, as is shown in the single-peak fit displayed in Fig. 2b. High resolution scans using Ge(111) as an analyzer indicate that the broadened peak has a double-peak structure, as can be seen in reciprocal space scans (see insets of Fig. 2c and 4c). The structural Bragg peaks exhibit no broadening or peak splitting, which suggests that there is a transition at T ≈ 10 K from a CM phase to an ICM phase. Note that the high resolution scans show double peaks in the L scans but not in the ω scans, and no other peaks were found in other scans along the high symmetry directions. This indicates that in the ICM phase, the magnetic propagation remains along the *c* axis, with a wavevector of (0 0 3/2 ± ε).

To obtain the width of the split peaks, we also fit the L scans using two Lorentzian-squared lineshapes with the same width, and the results are displayed in Fig. 2b. Note that below 15 K, unambiguous fits were impossible to obtain if both the width and intensities of the peaks were allowed to vary, and therefore the width was fixed to the value determined at 15 K. As can be seen in the figure, while the width from the single-peak fit increases monotonically above 10 K, the width of the peaks from the double-peak fit is roughly constant. This width is comparable to that of the (0 0 6) structural Bragg peak, which indicates that the Gd subsystem has long-range order below $T_N$. This observation is similar to the suggestion of simultaneous long-range ordering of the iron and rare-earth subsystems indicated by neutron scattering studies of Nd, Tb, and Ho ferroborates.[16-18]

The incommensurability, ε, calculated from the separation of the two peaks is shown in Fig. 2c. The onset of the ICM phase is determined as the temperature at which the splitting deviates from zero. In addition, it can be determined as the temperature at which the peak width obtained from a single-peak fit starts to increase, as can be seen in Fig. 2b. Both indicate that the onset of the splitting is ~10 K. As seen from Fig. 2c, ε increases continuously above 10 K, and reaches ~0.002 *c** near $T_N$. The small value of ε indicates a long-period ICM structure.

While Fig. 2a does not show an obvious change in the intensity of the (0 0 9/2) peak at $T_{SR}$, at which the magnetic moments are believed to reorient from the *c* axis to the basal plane, such changes are clearly seen at (0 0 3/2) and (0 0 15/2), as is shown in Fig. 3a. These changes can be explained by examining the resonant scattering cross-section for incident σ polarization, which for dipole transitions (i.e., 2p→5d) is proportional to:[29,30]

$$\left|\hat{M} \bullet \hat{x} \cos\theta - \hat{M} \bullet \hat{z} \sin\theta\right|^2 \qquad (1)$$

where **M** is the Gd magnetization, and the coordinate system shown in the inset to Fig. 1 is used. For resonant scattering at the Gd $L_{II}$ edge, θ is 8.84°, 27.64°, and 50.62° for (0 0 3/2), (0 0 9/2), and (0 0 15/2), respectively. Therefore as the temperature increases through $T_{SR}$, one expects the intensities of the (0 0 3/2) and (0 0 9/2) peaks to increase and that of the (0 0 15/2) peak to decrease— if the reorientation is from the *c* axis to the basal plane. The (0 0 3/2) and (0 0 15/2) peaks behave as expected, but the lack of an anomaly at (0 0 9/2) suggests that the reorientation at $T_{SR}$ is not a 90° rotation of the magnetization. Since nonresonant scattering data, which will be discussed in section III.B., indicate that the moments are along the *c* axis at T = 2 K, the behavior of the (0 0 9/2) peak intensity suggests that the moments are tilted out of the basal plane for $T > T_{SR}$.

To calculate the tilt angle, α, which we define as the angle between the Gd magnetization and the basal plane, we normalize the intensity of the (0 0 9/2) and (0 0 15/2) peaks to their values at T = 2 K, and introduce a parameter to account for the decrease in the Gd magnetization with increasing temperature. Since no azimuthal dependence was observed for the (0 0 9/2) peak,[31] which could be due to equal domain populations or a spiral structure in the basal plane, we average the scattering cross-section with respect to azimuthal angle, and calculate the tilt angles that are shown in Fig. 3b. A distinct transition characterized by a step-like change in the tilt angle can be seen, with the magnetization above $T_{SR} \approx 9$ K rotated ~45° out of the basal plane. We note that this reorientation temperature is close to $T_{SR}$ from previous reports.[3]

### B. ZERO-FIELD NONRESONANT X-RAY SCATTERING

Similar to Fig. 2, Fig. 4 shows the results for the nonresonant magnetic scattering at the (0 0 9/2) magnetic peak at 7 keV, which is well below the Gd $L_{III}$ edge (7.243 keV) and the Fe K edge (7.112 keV). Thus the peak intensity is from nonresonant scattering, with contributions from both the Fe and Gd moments. One sees from Fig. 4a that the variation of the integrated intensity is much different from that of the resonant scattering at the Gd $L_{II}$ edge. Starting from the lowest temperature, the intensity remains roughly constant until it increases suddenly at $T_{SR} \approx 9$ K, then it continues to increase and reaches a maximum at $T \approx 20$ K. It then decreases with increasing temperature and disappears at $T_N$. Fitting the data near $T_N$ with a power law as done for the resonant scattering data gives $T_N$ = 37.1 (±0.5) K, $\beta$ = 0.41(±0.07). $T_N$ is in good agreement with the value obtained from the resonant scattering data.

The jump in the intensity at $T_{SR} \approx 9$ K corresponds to the spin reorientation transition of the moments from the easy-axis state below $T_{SR}$ to the easy-plane state above $T_{SR}$. The nonresonant scattering intensity for incident $\sigma$ polarization is proportional to:[32]

$$\left|S_y \sin(2\theta)\right|^2 + 4\sin^4\theta \left|(S_x + L_x)\cos\theta + S_z \sin\theta\right|^2 \cos^2(2\theta_a) \qquad (2)$$

where S and L are the spin and orbital moments, $2\theta_a$ is the scattering angle of the analyzer, and the two terms are the $\sigma$ to $\sigma$ and $\sigma$ to $\pi$ components, respectively. We expect the spin moment to dominate since L = 0 for Gd ions and the orbital moment is generally quenched in Fe. Using (2), the scattering intensity of the (0 0 9/2) magnetic peak should increase by a factor of ~4 if the Fe and Gd moments are collinear. The observed factor of ~4 increase therefore suggests that the Fe moments are rotated 45° out of the basal plane just above $T_{SR}$, as are the Gd moments. This collinearity of the Fe and Gd subsystems is not maintained for increasing temperature, however, as indicated by the increase in intensity up to T ≈ 20 K. That is, the increasing intensity must arise either from an increase in the magnetic moment, or a decrease in the tilt angle. Since the magnetization does not increase with increasing temperature above T = 10 K,[6] the Fe moments must be rotating further toward the basal plane as the temperature increases. Changes in the energy gap and line width near T = 20 K have been discovered in antiferromagnetic resonance measurements,[8] and our x-ray scattering results suggest that a change in the tilt angle of the Fe subsystem is their possible origin. The collinearity near $T_{SR}$ may be due to an increase in the interaction between the Fe and Gd subsystems as the Gd magnetization increases with decreasing temperature, and the interaction presumably drives the spin reorientation at $T_{SR}$ when it is sufficiently strong. We note that a similar explanation for the spin reorientation in Ho ferroborate has recently been proposed.[18]

At the (0 0 3/2) magnetic peak, (2) indicates that the scattering intensity will increase by more than two orders of magnitude when the magnetization reorients from the *c* axis to the easy-plane state with a tilt angle of 45° for both subsystems. The peak intensity of the (0 0 3/2) above $T_{SR}$ is ~100/s with a background of ~3/s, thus the peak will be too weak to observe if the moments are along the *c* axis below $T_{SR}$. Indeed, no (0 0 3/2) peak was observed at T = 2 K. This lends support to our assumption that the magnetization is along the *c* axis at T = 2 K, which was used to calculate the tilt angles reported in section III.A.

Figs. 4b and c show the (0 0 9/2) peak width and the incommensurability as functions of temperature. The onset of the ICM phase (dashed line) occurs at approximately the same temperature as observed using resonant x-ray scattering, i.e., ~1 K above the intensity jump at $T_{SR}$, shown in Fig. 4a (dotted line). We note that this onset temperature is very close to $T_M$, which is 0.7 K above $T_{SR}$ as found in dielectric measurements.[3] This suggests that $T_M$ is likely the CM-ICM phase transition temperature, and thus the easy-plane, AFM phase can be further divided into two phases: CM AFM for $T_{SR} < T < T_M$, and ICM AFM for $T > T_M$.

## C. X-RAY SCATTERING IN A MAGNETIC FIELD

In order to further understand the FIP phase in $GdFe_3(BO_3)_4$, we performed magnetic x-ray scattering measurements with the sample in a magnetic field applied along the *a* axis. Fig. 5 shows the field dependence of the (0 0 9/2) magnetic peak intensity at the Gd $L_{II}$ edge normalized to the intensity of the (0 0 3) structural Bragg peak, and the L scan peak width, at T = 8 and 25 K. Also indicated by the two vertical lines in the figure are the critical fields for the FIP phase at the two temperatures, as determined by previous work.[2]

One sees from Fig. 5a that at T = 8 K there is an increase in the intensity for B > 2 T, which indicates the reorientation of the Gd moments, as will be discussed below. The peak width remains roughly constant for B ≤ 2.5 T, and then decreases for stronger applied field. Note that the field required to reorient the spins agrees reasonably well with the critical field for the FIP phase. At T = 25 K, the intensity increases only slightly in the low field region, near the 0.5 T critical field for the FIP phase, and the peak width decreases gradually with increasing field. The small change in the intensity is consistent with the fact that there is no spin reorientation because the spins are in the easy-plane state at this temperature in zero field.

For the horizontal scattering geometry at the X21 beamline shown in the inset to Fig. 5a, the incident beam is π dominated, and the resonant scattering intensity is proportional to:[29,30]

$$\left|\hat{M}\bullet\hat{x}\cos\theta+\hat{M}\bullet\hat{z}\sin\theta\right|^2+\left|\hat{M}\bullet\hat{y}\sin(2\theta)\right|^2\cos^2(2\theta_a) \qquad (3)$$

where the first term is the π to σ component, and the second term is the π to π component. Note that there is an extra term for unrotated polarization compared to scattering with incident σ polarization, and it disappears when the magnetization is in the scattering plane. From (3), as the Gd moments reorient from the *c* axis toward the basal plane at T = 8 K, the intensity will increase, as observed. At T = 25 K, the increase in the intensity of the peak is small, and can be explained by the rotation of the Gd moments away from the direction of the applied field, and therefore into the scattering plane. The field dependences at both temperatures indicate that the Gd moments are rotated toward the basal plane in the FIP phase.

Fig. 6 shows the field dependence of the (0 0 9/2) magnetic peak intensity and its width at T = 8 and 25 K, taken off resonance at 7 keV, and with the intensity normalized to the (0 0 3) structural Bragg peak intensity. The peak intensity and width decrease dramatically for T = 8 K at ~2 T. At T = 25 K, the intensity and peak width decrease monotonically until they reach a minimum at B ≈ 0.8 T. In the FIP phase, both the peak intensity and width are similar at the two temperatures. There are two possible reasons for the decrease in the peak width in the applied magnetic field. The first is the coalescence of magnetic domains, which may be caused by the moments rotating to the plane that is perpendicular to the magnetic field. The second reason is that the magnetic field drives the moments from the ICM phase to the CM phase, which eliminates the peak splitting. We were unable to resolve the two peaks in the ICM phase in zero field because of the small separation of the two peaks and the relatively low resolution due to the larger divergence of the incident beam in the horizontal plane. However, the peak width in the FIP phase is significantly smaller, as shown in Fig. 6b, and comparable to that of the CM phase in Fig. 4b, which suggests that the magnetic structure in the FIP phase is commensurate.

The intensity decrease at both temperatures in the FIP phase is opposite to what was observed for the resonant scattering. For a π incident beam, the nonresonant scattering intensity is proportional to:[32]

$$4\sin^4\theta\left|(S_x+L_x)\cos\theta-S_z\sin\theta\right|^2+\sin^2(2\theta)\left|S_y+2L_y\sin^2\theta\right|^2\cos^2(2\theta_a) \quad (4)$$

where the first and second terms are the π to σ and π to π components, respectively. Using (4), one calculates an increase in the intensity of the (0 0 9/2) magnetic peak when the moments reorient toward the basal plane, which is expected at T = 8 K and B = 2 T. At T = 25 K, the moments are in the easy-plane state without the applied field. If the moments rotate to the scattering plane in the field without changing the tilt angle, one expects that the intensity will decrease, but merely by ~40%. The observed intensities at both temperatures in the FIP phase are therefore inconsistent with our expectations based on the resonant scattering data. A possible scenario for the significant decrease in the scattering intensity at both temperatures is that the Fe moments are canted so that they acquire a ferromagnetic component in the magnetic field, and the ferromagnetic component does not contribute to the (0 0 9/2) magnetic peak intensity.

The temperature dependences of the intensity of the (0 0 9/2) magnetic peak normalized to the (0 0 3) structural Bragg peak at the Gd $L_{II}$ edge and 7 keV are shown in Fig. 7. For the resonant scattering (Fig. 7a), there is an increase in the intensity between T = 8 and 10 K for both B = 0 and 0.2 T, which indicates the spin reorientation. Note that an increase in the (0 0 9/2) intensity for B = 0 T is seen here but not in Fig. 2a because of the different scattering geometry. Within the FIP phase the intensity is enhanced slightly, which is likely due to the Gd moments rotating away from the direction of the applied field toward the scattering plane. For T > 20 K, for which the FIP phase is not observed for both B = 0 and 0.2 T, the two curves match almost perfectly.

The nonresonant scattering data behave quite differently, as shown in Fig. 7b. Below $T_{SR} \approx 9$ K, the moments are in the easy-axis state, and the intensity at the two fields is similar. However above $T_{SR}$ and inside the FIP phase (i.e., for B = 0.2 T), the intensity is much smaller than at B = 0 T, which is consistent with the data in Fig. 6. For B = 0 T the intensity first increases and reaches a maximum at ~20 K, then decreases gradually, similar to the data shown in Fig. 4a. In contrast for B = 0.2 T, the intensity decreases above $T_{SR}$ and reaches a *minimum* at $T \approx 20$ K. That the magnetic peak intensity at B = 0.2 T reaches a minimum at the same temperature as it reaches a maximum at B = 0 T suggests that the Fe moment component in the basal plane becomes ferromagnetic in the FIP phase.

## IV. DISCUSSION

Our nonresonant x-ray scattering in a magnetic field data suggest that there is a CM to ICM transition at $T_M$, and it has been established that field-induced polarization occurs below $T_M$.[3] Combining our results with those in previous work,[2] we conclude that the FIP phase occurs only in a CM structure with the Fe and Gd moments tilted toward the basal plane. Since a field **H**||***c*** lowers $T_{SR}$ and does not change $T_M$,[3] polarization occurs below 9 K only when H is strong enough to reorient the moments toward the basal plane. In contrast, a low field **H**||***a*** increases $T_M$ but does not change $T_{SR}$.[3] Below $T_{SR}$, a strong field is needed to reorient moments from the *c* axis toward the basal plane to allow the polarization. Above $T_{SR}$, however, the moments are tilted toward the basal plane, and only a relatively weak field is needed to drive them from the ICM to the CM phase to allow the polarization. Therefore, our results are consistent with previous works, and also provide insight into previous observations.

The onset of ferroelectricity coinciding with the ICM-CM transition is also found in $RMn_2O_5$ (R=Ho, Er, Y, Bi).[33] Betouras *et al.*[15] proposed a theory to explain the correlation between a CM structure and ferroelectricity. In that theory, the direction of polarization is required to be parallel to the magnetic propagation vector $\mathbf{Q_m}$. For $GdFe_3(BO_3)_4$, however, the ferroelectricity is found to be parallel to the applied field while the magnetic propagation vector $\mathbf{Q_m}$ is along the *c* axis. Hence, this mechanism can't explain the field-induced polarization when **H**||***a***.

Zvezdin *et al.*[34] proposed a macroscopic theory to explain the multiferroicity in $GdFe_3(BO_3)_4$. They point out that as the moments rotate toward the basal plane, the symmetry of the crystal lowers from trigonal to monoclinic if the moments are parallel to one of the second-order axes, and becomes triclinic if the moments deviate from the second-order axes in the basal plane. This symmetry breaking through magnetic ordering allows the spontaneous electric polarization. Although the theory correctly describes many aspects of the multiferroic behavior of the material, it can't explain why a CM structure is necessary for ferroelectricity because the detailed magnetic structure is not needed in this theory.

It is known that ferroelectricity in $GdFe_3(BO_3)_4$ correlates with the magnetostriction.[34] A common origin of the ferroelectricity is the relative displacement of the oxygen ion from the transition metal ions caused by a distortion. Sergienko and Dagotto proposed a Hamiltonian based on the DM interaction:[35]

$$H_{DM}(\mathbf{r}_n) = \sum_n D(\mathbf{r}_n) \cdot [\mathbf{S}_n \times \mathbf{S}_{n+1}] + H_{el} \qquad (5)$$

where $r_n = (-1)^n r_0 + \delta\mathbf{r}_n$, $r_0$ is the orthorhombic distortion, $\delta\mathbf{r}_n$ is the further distortion associated with the ferroelectricity, $H_{el} = 1/2(\kappa_x x_n^2 + \kappa_y y_n^2 + \kappa_z z_n^2)$ is the elastic energy gain from the distortion, and $D(\mathbf{r}_n)$ is the Dzyaloshinskii-Moriya vector. Here the coordinate system is set up so that the *x* axis is along the crystallographic *a* axis (parallel to the applied magnetic field), the *z* axis is along the *c* axis, and the *y* axis is parallel to $\mathbf{b}^*$. We assume that the FIP phase is a canted AFM phase, which is relatively common in the trigonal crystal,[36] and

$$\mathbf{S}_n = \mathbf{S}_c + \mathbf{S}_0 \cos(n\, q_m d) \qquad (6)$$

where Sc and $S_0$ are the ferromagnetic and antiferromagnetic components of the Fe spin magnetization, respectively; $\mathbf{q_m}$ is the magnetic propagation vector; and d is the distance between the neighboring layers of Fe spins along the *z* axis. For simplicity, here we only consider the Fe spins, which have the dominant contribution to the magnetization of $GdFe_3(BO_3)_4$.

In the FIP phase, the component parallel to the basal plane becomes ferromagnetic and opposite to the magnetic field to minimize the Zeeman energy, thus $\mathbf{S}_n = -S_c\,\mathbf{e_x} + S_0 \cos(n\, q_m d)\,\mathbf{e}_z$. In reference 35, a **D** vector with linear dependence on the coordinates was proposed to explain the ferroelectricity occuring in an ICM phase. We show here that a **D** vector with a different form is possible to explain the ferroelectricity in a canted CM phase: for example, $D(\mathbf{r_n}) = \gamma x_n^{\mathbf{2}} \mathbf{e_y}$ (or $D(r_n) = \gamma x_n^{2j} e_y$, j = nonzero integer, gives a similar result). Expanding the Hamiltonian and only keeping the leading term of $\delta\mathbf{r}_n$ in $D(\mathbf{r_n})$ and $H_{el}$, the portion of the Hamiltonian that depends on $\delta\mathbf{r}_n$ is

$$H_{DM}(\delta \mathbf{r}_n) = 4\gamma \sum_n (-1)^n x_0 \delta x_n S_c S_0 \sin(n+\frac{1}{2}) q_m d \sin\frac{1}{2} q_m d$$

$$+1/2 \sum_n (\kappa_x \delta x_n^2 + \kappa_y \delta y_n^2 + \kappa_z \delta z_n^2) \quad (7)$$

Minimizing $H_{DM}(\delta \mathbf{r}_n)$ with respect to $\delta \mathbf{r}_n$, we obtain

$$\delta x_n = (-1)^{n+1} \frac{4\gamma x_0}{\kappa_x} S_c S_0 \sin(n+\frac{1}{2}) q_m d \sin(\frac{1}{2} q_m d) \quad (8)$$

and $\delta y_n = \delta z_n = 0$.

For $GdFe_3(BO_3)_4$, $q_m d = \pi$, and (8) reduces to

$$\delta x_n = -\frac{4\gamma x_0}{\kappa_x} S_c S_0 \quad (9)$$

which does not depend on n, and thus leads to a net polarization along the *x* axis. For an ICM phase, however, $\delta x_n$ depends on n, and its sum vanishes— thus there is no net polarization. In the general case, one can see from (8) that if the wavevector of the distortion in $\mathbf{r_0}$ is equal to $q_m$, $\delta x_n$ will not depend on n and there will be a net polarization. This is possible in the CM phase but very unlikely in the ICM phase, because the underlying physics of the $\mathbf{r_0}$ and $\delta x_n$ distortions is different.

## V. CONCLUSIONS

In summary, we have used x-ray scattering techniques to study the magnetic structure of $GdFe_3(BO_3)_4$, both with and without applied magnetic field. Resonant scattering data show unambiguously that the Gd moments order at $T_N$. In zero field, we verified that both the Gd and the Fe moments have a reorientation transition at $T_{SR} \approx 9$ K. However, above $T_{SR}$ the Gd and the Fe moments behave quite differently: both the Gd and Fe moments are tilted ~45° away from the basal plane just above $T_{SR}$; then while the Gd tilt angle is essentially unchanged with increasing temperature, the Fe moments tilt further toward the basal plane. Using a high resolution analyzer, we discovered that ~1 K above the spin reorientation transition there is another transition from a CM to a long-period ICM structure. The incommensurability increases monotonically and reaches ~0.002 *c** near $T_N$.

For a magnetic field applied along the *a* axis, x-ray scattering confirms that strong magnetic field can drive the moments to reorient below 9 K. The decrease in the magnetic peak width at T = 25 K as the FIP phase is entered supports the conclusion that the magnetic structure is commensurate in the FIP phase. The onset of the FIP phase from our data is in good agreement with previous work. The resonant x-ray scattering indicates that in the FIP phase, the Gd moments are tilted toward the basal plane and in the plane perpendicular to the magnetic field. Therefore the magnetic structure of the Gd subsystem in the FIP phase is similar to its structure in zero field between 9 and 10 K. The nonresonant x-ray scattering suggests that the Fe moments also rotate into the plane perpendicular to the field, and are canted with the ferromagnetic component in the basal plane. Combining our results with previously published results, we conclude that commensurability and moments lying close to the basal plane are necessary for field-induced polarization in this compound. We propose a DM interaction Hamiltonian to explain the correlation between the CM structure and the FIP. According to this Hamiltonian, spin canting and the orthorhombic distortion play important roles in the correlation between the ferroelectricity and a CM structure.

## ACKNOWLEDGMENTS

We thank W. Schoenig, D. Coburn, S. Wilkins, and J. Hill for their support at X22C, S. LaMarra for his support at X21, and D. Connor for a critical reading of the manuscript. Use of the National Synchrotron Light Source, Brookhaven National Laboratory, was supported by the US Department of Energy, Office of Science, Office of Basic Energy Sciences, under Contract No. DE-AC02-98CH10886.

Figure Captions

Fig. 1. (Color online) Energy dependence of peak intensities at the Gd (a) $L_{II}$ and (b) $L_{III}$ edges. Inset in (a) shows the π (●) and σ (Δ) components of the (0 0 3/2) magnetic peak intensity. Inset in (b) is a sketch of the scattering geometry on beamline X22C.

Fig. 2. (Color online) Temperature dependence of the resonant scattering of the (0 0 9/2) magnetic peak at the Gd $L_{II}$ edge. (a) The integrated intensity normalized to that of the (0 0 6) structural Bragg peak. Inset shows a power law fit near $T_N$. (b) The width of the L scans from single- and double-peak fits. (c) The incommensurability. Inset shows the splitting of the peak in an L scan measured at T = 35.5 K.

Fig. 3. (Color online) (a) The resonant scattering, integrated intensities of the (0 0 3/2) (●) and (0 0 15/2) (○) magnetic peaks normalized to that of the (0 0 3) structural Bragg peak, at the Gd $L_{II}$ edge. (b) The tilt angle of the Gd moments with respect to the basal plane.

Fig. 4. (Color online) Temperature dependence of the nonresonant scattering of the (0 0 9/2) magnetic peak at 7 keV. The dotted line indicates the temperature of the spin reorientation, and the dashed line indicates the temperature of the CM-ICM transition. (a) The integrated intensity normalized to that of the (0 0 6) structural Bragg peak. Inset shows a power law fit near $T_N$. (b) The width of the L scans from single- and double-peak fits. (c) The incommensurability. Inset shows the splitting of the peak in an L scan measured at T = 22.5 K.

Fig. 5. (Color online) Field dependence of the resonant scattering of the (0 0 9/2) magnetic peak at the Gd $L_{II}$ edge at T = 8 K (○) and 25 K (Δ). The vertical lines indicate the critical field for the FIP phase at the two temperatures. (a) Integrated intensity normalized to that of the (0 0 3) structural Bragg peak. Inset shows a schematic diagram of the X21 scattering geometry. (b) Peak width of L scans.

Fig. 6. (Color online) Field dependence of the nonresonant scattering of the (0 0 9/2) magnetic peak at 7 keV at T = 8 K (○) and 25 K (Δ). The vertical lines indicate the critical field for the FIP phase at the two temperatures. (a) Integrated intensity normalized to that of the (0 0 3) structural Bragg peak. (b) Peak width of L scans.

Fig. 7. (Color online) Temperature dependence of the integrated intensity of the (0 0 9/2) magnetic peak normalized to that of the (0 0 3) structural Bragg peak at B = 0 T (●) and 0.2 T (○). The vertical lines are the temperature boundaries of the FIP phase for B = 0.2 T. (a) Resonant scattering at the Gd $L_{II}$ edge. (b) Nonresonant scattering at 7 keV.

Fig.1

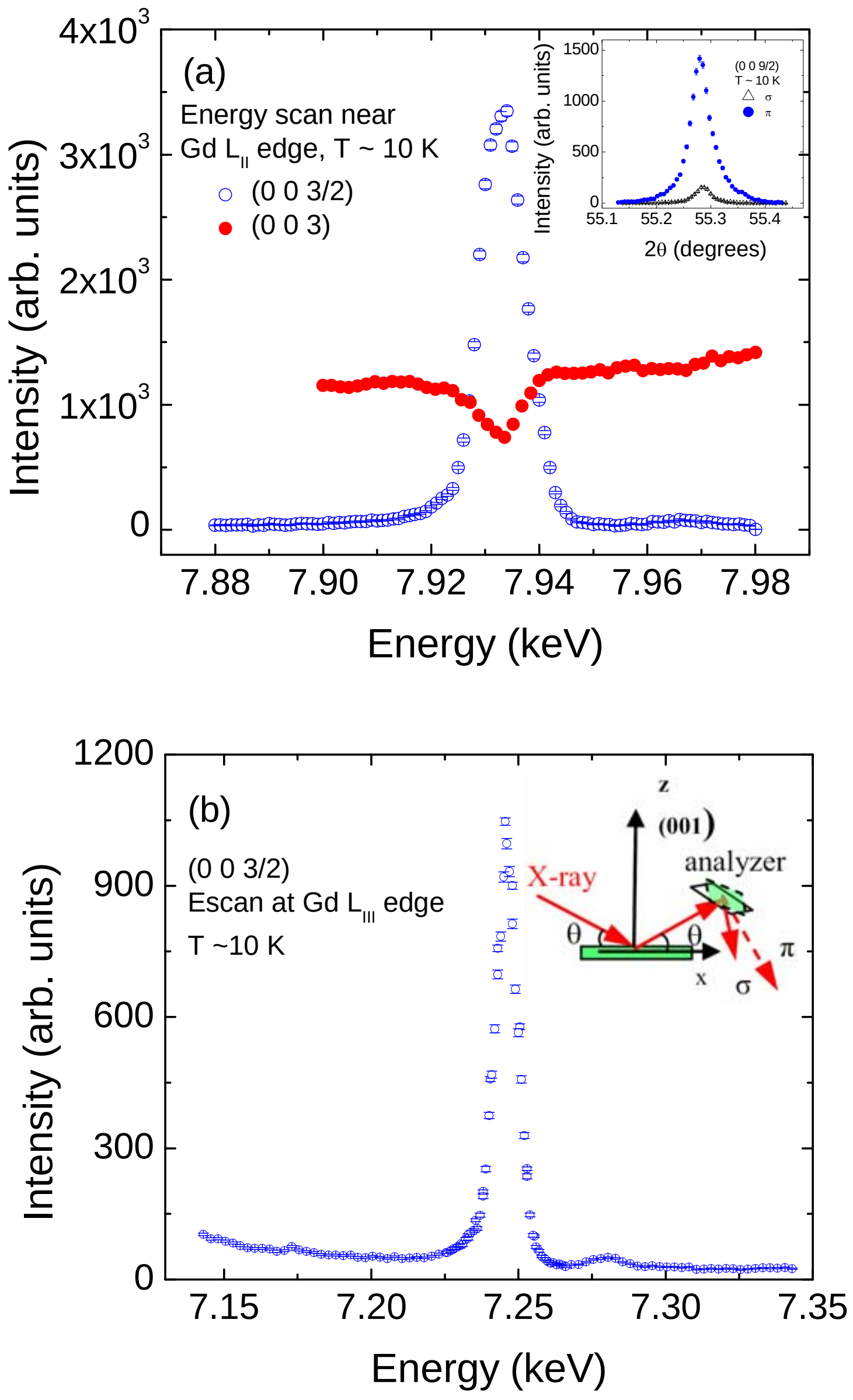

Fig. 2

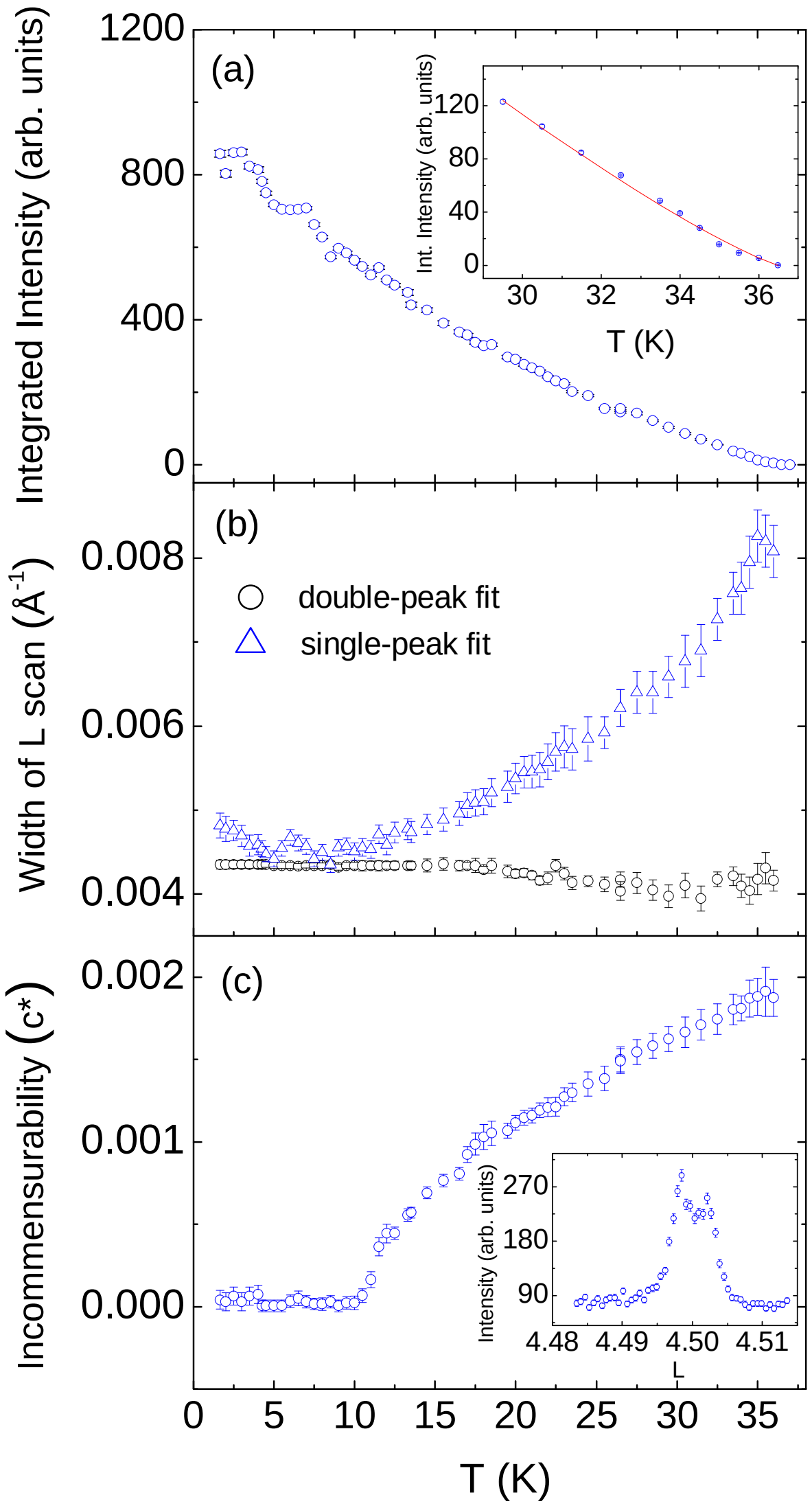

Fig. 3

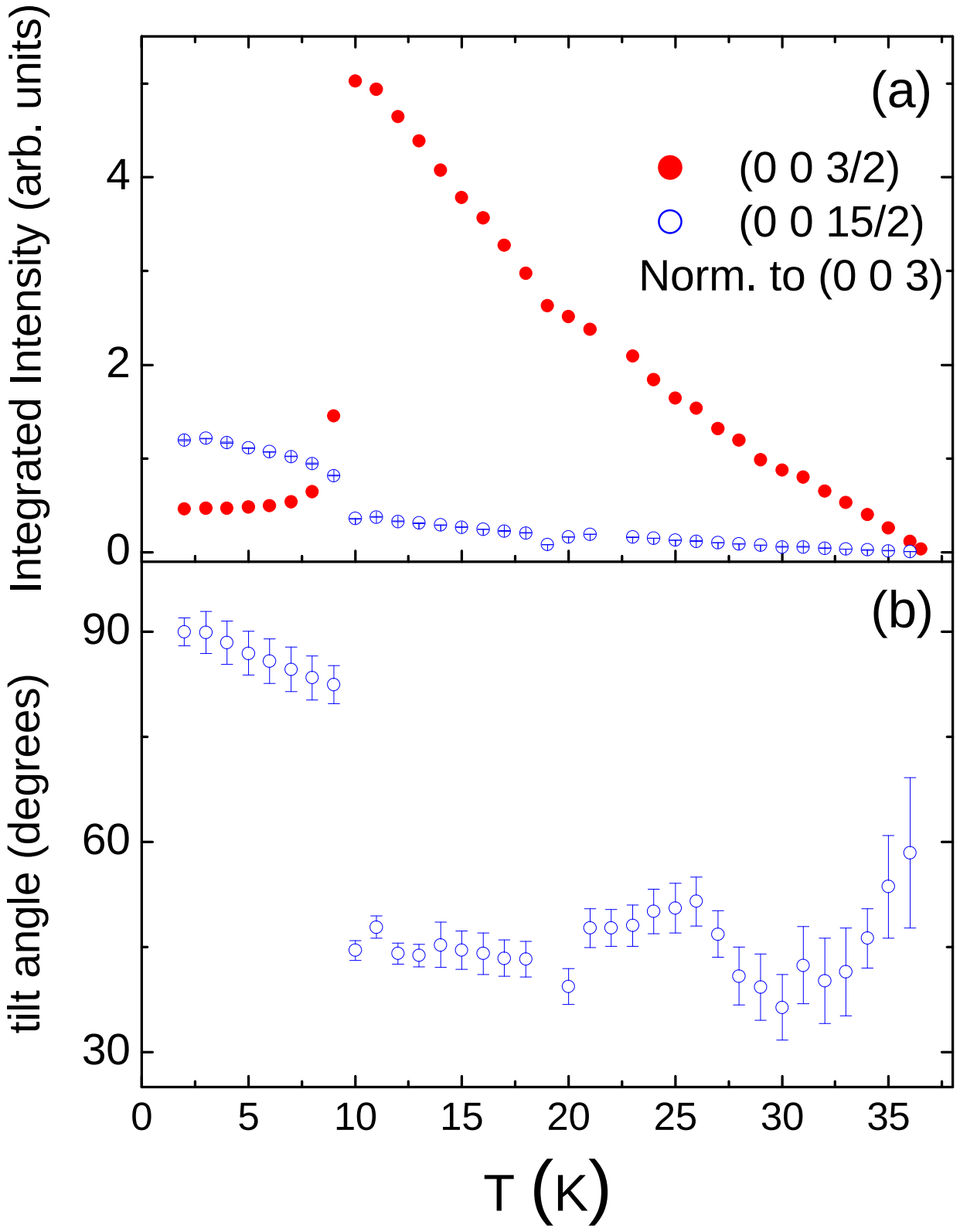

Fig. 4

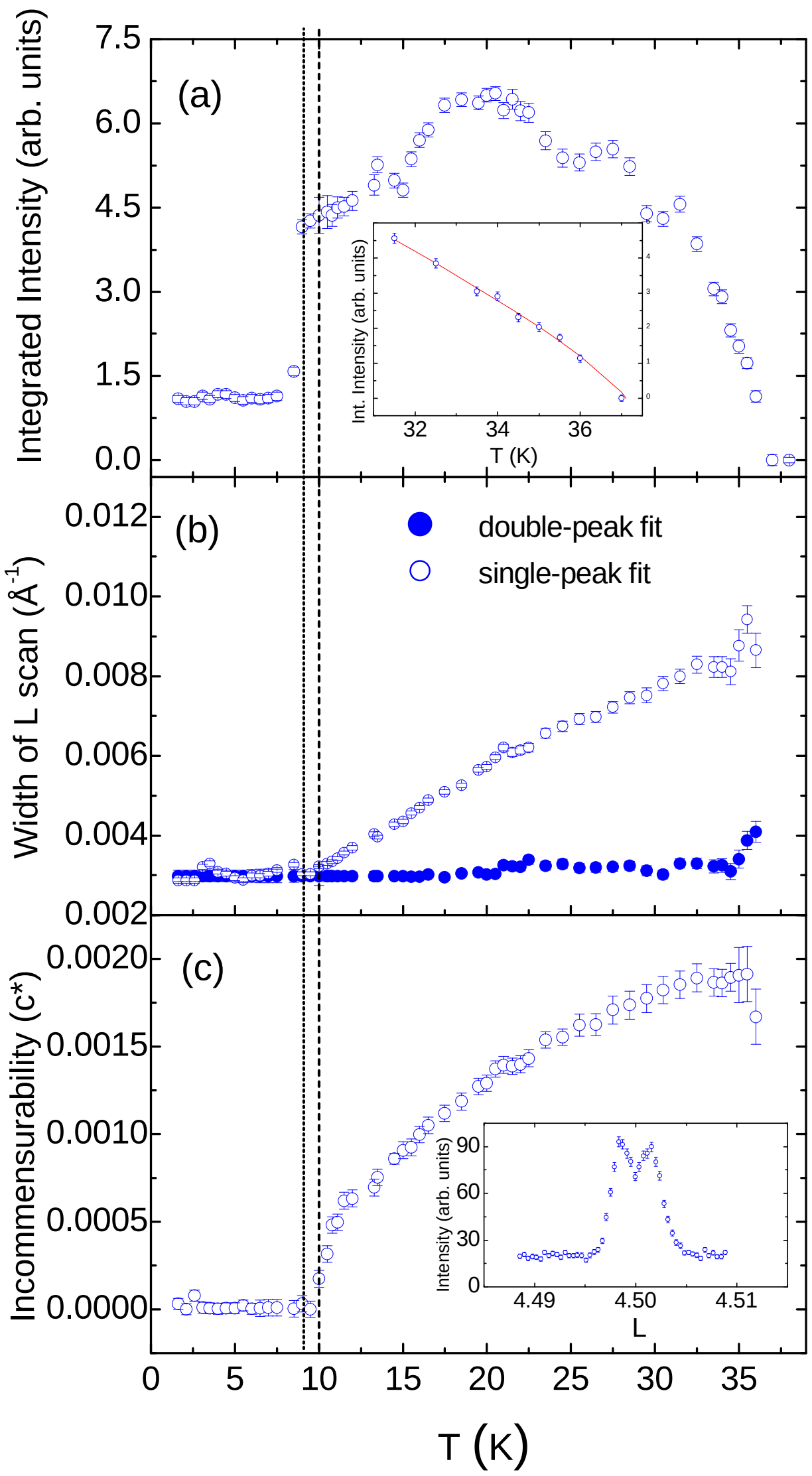

(a)
(b)
(c)
Integrated Intensity (arb. units)
Width of L scan (Å$^{-1}$)
Incommensurability (c*)
double-peak fit
single-peak fit
Int. Intensity (arb. units)
T (K)
Intensity (arb. units)
L
T (K)

Fig. 5

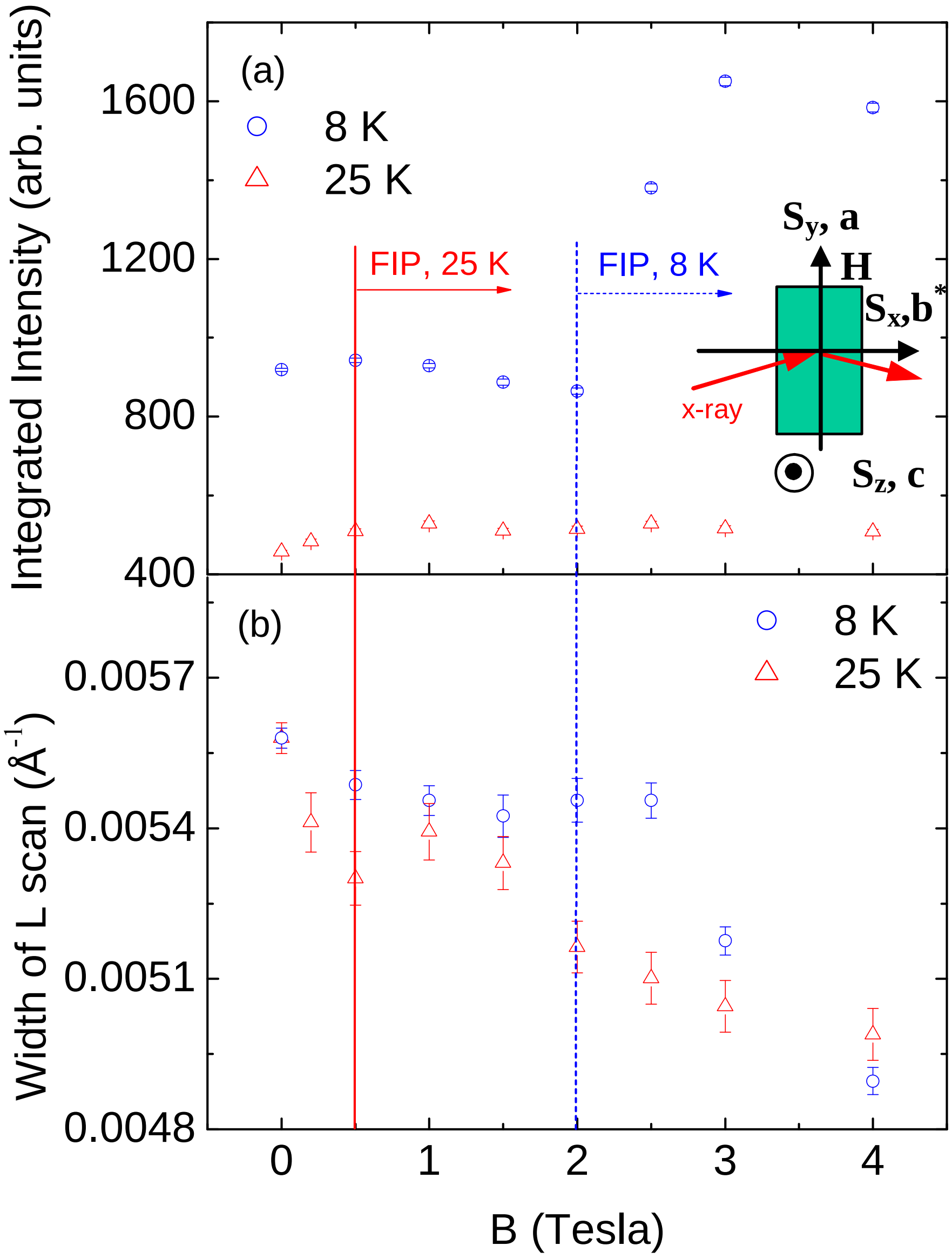

Fig. 6

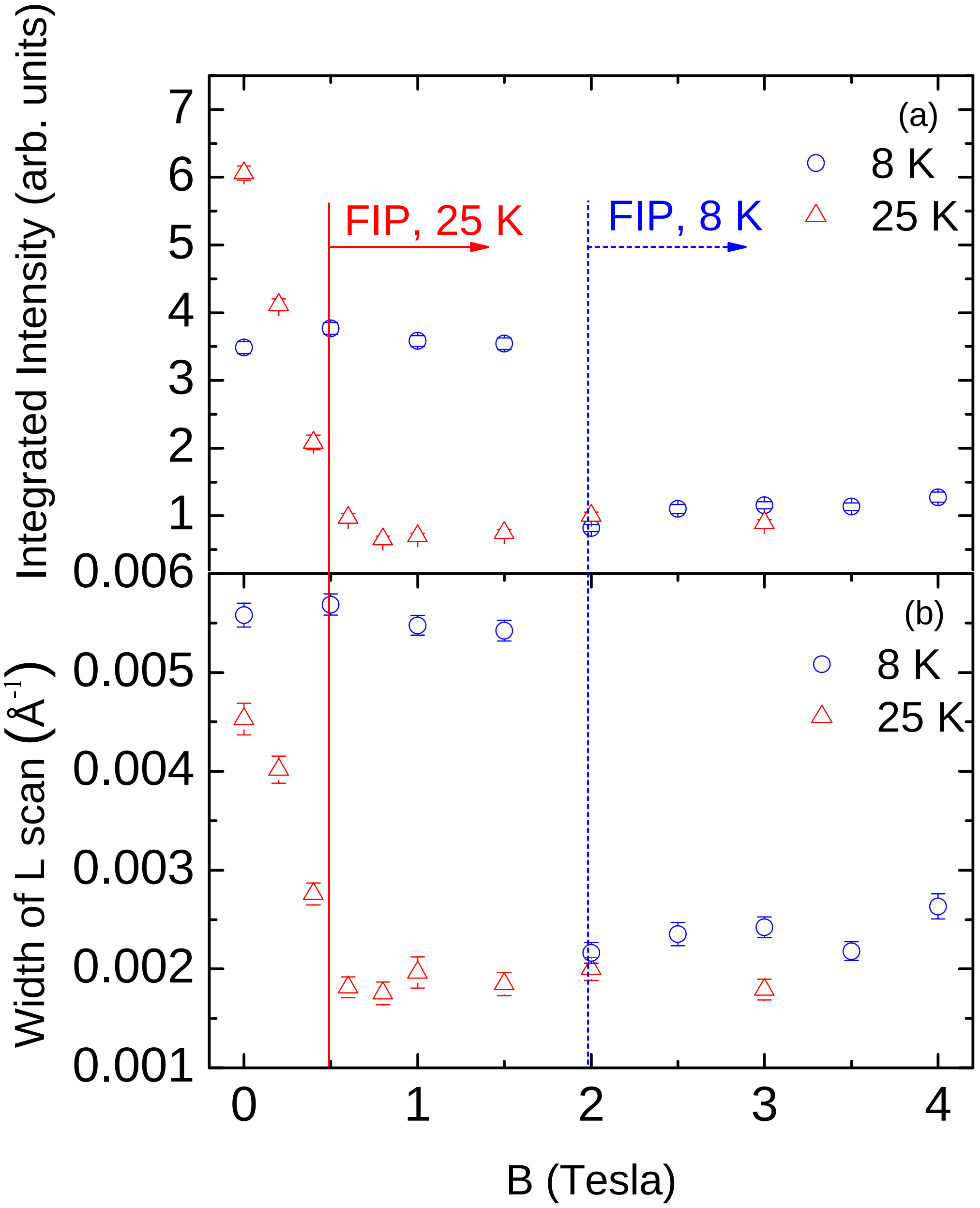

Fig. 7

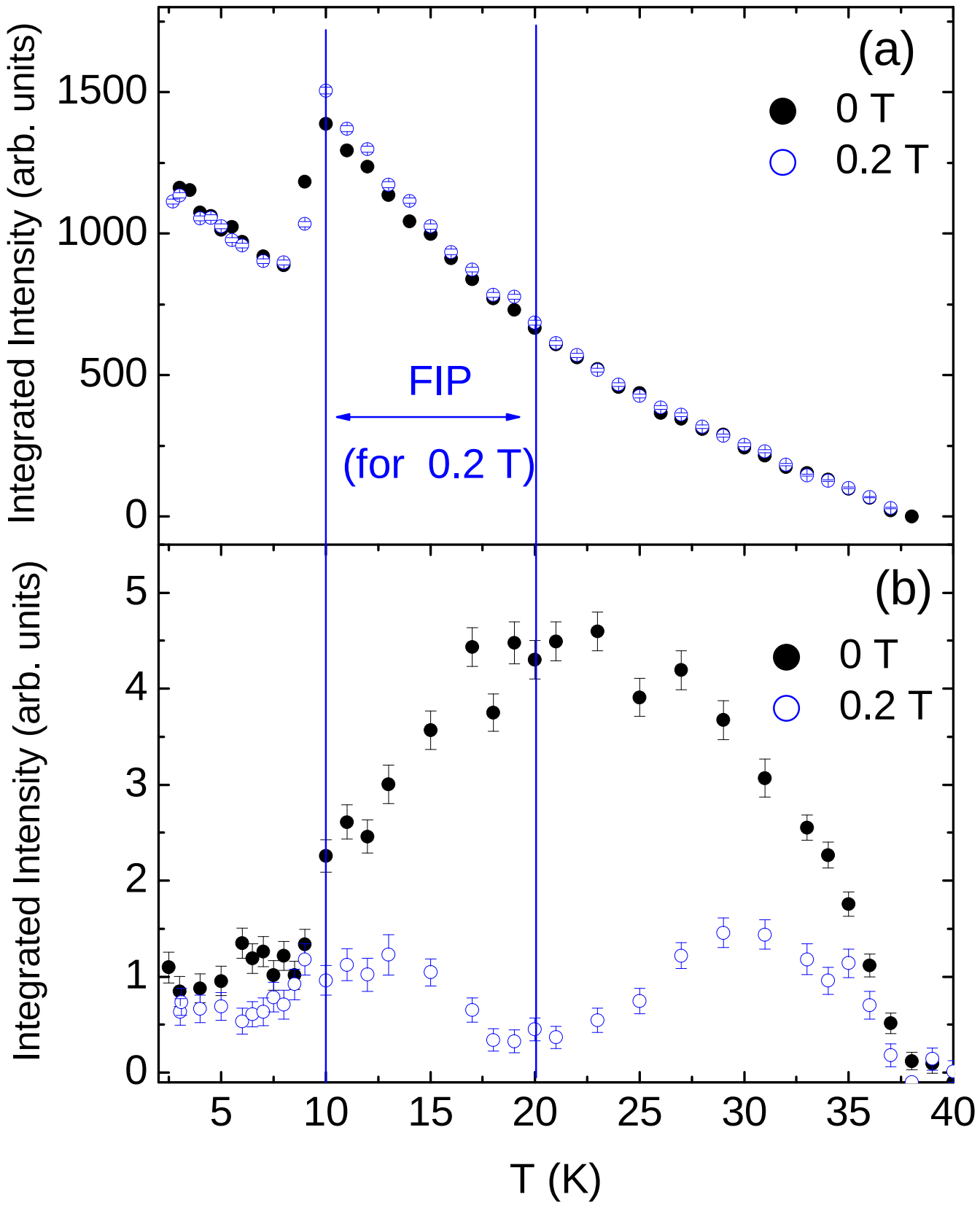